\documentclass[aps,prl,twocolumn]{revtex4}

\usepackage{graphicx}
\usepackage{bm}
\usepackage{epstopdf}

\begin{document}

\title{Hopping Conduction and Bacteria: Transport in Disordered Reaction-Diffusion Systems}
\author{Andrew R. Missel}
\email[]{missel@uiuc.edu}
\author{Karin A. Dahmen}
\affiliation{Physics Department, University of Illinois at Urbana-Champaign, Urbana, IL 61801, USA}

\date{\today}

\begin{abstract}
We report some basic results regarding transport in disordered reaction-diffusion systems with birth ($A\to2A$), death ($A\to0$), and binary competition ($2A\to A$) processes.  We consider a model in which the growth process is only allowed to take place in certain areas---``oases''---while the rest of space---the ``desert''---is hostile to growth.  In the limit of low oasis density, transport is mediated through rare ``hopping'' events, necessitating the inclusion of discreteness effects in the model.  By first considering transport between two oases, we are able to derive an approximate expression for the average time taken for a population to traverse a disordered medium.
\end{abstract}

\maketitle

Reaction-diffusion (RD) models are often used to study population dynamics, animal coat patterns, and a great variety of other systems, both biological and non-biological \cite{murray, isingglauber, kroon1993, birch2007bbf}.  The effects of quenched spatial disorder in the reaction rates on the behavior of these models have been difficult to discern---in some cases, a renormalization group analysis yields runaway flows \cite{runawayRG}---but some progress has been made \cite{MCofcontactproc, disorderPTVojta, jaewookcontact,parkanddeem,birch2007bbf}.  In particular, Nelson and coworkers have looked at the effects of convection and quenched spatial disorder on the evolution of a population density described by a generalization of the Fisher/KPP equation given by
\begin{equation}
\frac{\partial c(\bm{x},t)}{\partial t}=D\nabla^{2}c(\bm{x},t)+U(\bm{x})c(\bm{x},t)-qc(\bm{x},t)^{2},\label{fishnels}
\end{equation}
where $c(\bm{x},t)$ represents the population density, $D$ is a spatially homogenous diffusion constant, $U(\bm{x})$ is a spatially inhomogeneous growth term, and $q=b{\ell_{0}}^{d}$ is a competition term ($b$ is a competition rate and $\ell_{0}$ is the microscopic length scale at which two particles will compete with one another) \cite{nelsonshnerb1998}.  One simple form of inhomogeneity considered in these works is a ``square well'' potential $U(\bm{x})$ which consists of a uniform space with negative growth rate---termed the ``desert''---in which a single region of positive growth rate---an ``oasis''---is placed.  This model has proven to be applicable to experiments with bacteria populations in adverse environments \cite{exper2}.

Other important studies have focused on the effects of spatial inhomogeneities in both reaction and diffusion rates on the speed and roughness of front solutions to reaction-diffusion equations \cite{fisherwavesrandommedia,flucsinfronts}.  These works consider a parameter regime---which we dub \emph{fertile}---in which a front solution exists in the absence of disorder (for (\ref{fishnels}), this regime corresponds to $U(x)=U>0$).  The nature of transport in the opposite parameter regime in which the average growth rate is negative---which we dub the \emph{hostile} regime---has not been studied as much.  In this regime, there may be localized stationary solutions centered around regions of positive growth rate---oases \cite{nelsonshnerb1998}---as well as noise-induced front solutions for sufficiently large variations in growth rate \cite{noisefronts}.

In this letter, we report some results on the nature of transport in a hostile disordered system; specifically, we consider the case of a desert into which identical oases are placed randomly at low density.  Because transport between oases involves the movement of a low population density, it is natural to assume that discreteness effects---fluctuations about the mean field theory---might come into play.  For this reason, we consider the stochastic dynamics underlying the differential equation (\ref{fishnels}) in which discrete particles are allowed to diffuse, reproduce ($A\to2A$), die ($A\to0$), and compete ($2A\to A$).  Using a novel method, we find the probability density function for the time at which a particle from a populated oasis first reaches an unpopulated oasis---the \emph{first passage time}; we then show how, for sufficiently low oasis density, this result can be used along with the theory of hopping conduction to estimate the mean transit time across the system.

We start by studying transport between two oases, one of which---we will call this the first oasis---is initially populated.  We define the infection time to be the time elapsed before the second oasis reaches the population level at which the first oasis started.  This time can be broken into two parts: the time $T_{\text{transit}}$ that it takes the population to reach the second oasis and the time $T_{\text{growth}}$ it takes for the population at the second oasis to grow to the specified level.  We assume very fertile oases; that is, we assume that the particles which first reach the second oasis will reproduce immediately, ignoring the possibility that they may return to the desert and die.  (This can be accomplished by having oases with high growth rates or by ``seeding'' the oases with a second species $B$ and including a new very fast reaction $A+B\to2A$.)  This makes $T_{\text{transit}}$ identical to the first passage time (FPT) of the process---that is, the time it takes for the first particle to reach the second oasis.  We will calculate the probability distribution of this time---the first passage time probability distribution function (FPT PDF).

Consider two oases of radius $a$ in $d$ dimensions whose centers are separated by a distance $R$.  The oases each have growth rate $y$, and the desert between them has death rate $z$.  The competition rate is $b{\ell_{0}}^{d}$, where $\ell_{0}$ is a microscopic length scale.  If $R$ is sufficiently small, we expect that the FPT should be linear in $R$: a ``wavefront'' of particles radiates out from the first oasis due to diffusion, its amplitude decreasing due to the death term, eventually reaching a level small enough that mean field theory is invalid.  The distance $R_{\text{lin}}$ at which this linear transport breaks down, then, should be approximately the same as the distance at which mean field theory breaks down.  This distance can be obtained by solving the mean field differential equation (\ref{fishnels}) and looking at the solution for large times; the distance where the population density falls off to $\sim{\ell_{0}}^{-d}$ is the distance we are looking for.  In one dimension, $c(x)\simeq c_{0}\,e^{-\sqrt{z/D}\,x}$ for large $x$.  The constant $c_{0}$ is of order $y/(b{\ell_{0}}^d)$ for small values of $y$, and thus an estimate for $R_{\text{lin}}$ is $\sqrt{D/z}\,\ln(y/b)$.  In higher dimensions, the relevant length scale is smaller, but we will use the above expression as a rough estimate.

Because the competition process $2A\to A$ is not a significant mechanism of particle destruction for distances beyond $R_{\text{lin}}$, the particles' interactions with each other can be ignored.  With this simplification in mind, a model for the behavior of the system at large oasis separation can be formulated by replacing the first oasis with desert and a point source ($0\to A$) which creates $N$ \emph{non-interacting} particles per unit time.  If the creation rate is chosen so that the average flux of particles at $R_{\text{lin}}$ is the same as for the model with competition, the FPT statistics should be nearly identical for large oasis separations.  We will refer to this simplified model as the linear model with a source, and to the full model with competition as the nonlinear model.

Since the particles are non-interacting in the linear model with a source, the full FPT PDF, which we will denote $f_{N}(R,t)$, can be written in terms of the one-particle FPT PDF $f_{1}(R,t)$.  We will do this as follows: let the source be at the origin, and define $S(R,t)=1-\int_{0}^{t}dt'\,f_{1}(R,t')=1-P_{\text{hit}}(R,t)$ to be the probability that a particular particle has not reached the oasis located a distance $R$ away by time $t$.  The probability $P_{\text{none}}(R, t)$ that this site has \emph{never} been visited by \emph{any} particle by time $t$ is just a product of the probabilities $S(R, t)$ that each of the particles has never visited it:
\begin{equation}
P_{\text{none}}(R, t)\,=\prod_{\tau=0, \Delta t, \ldots}^{t}[S(R,\tau)]^{N}.\label{probnone1}
\end{equation}
Taking the natural logarithm, letting $\Delta t\to0$ with $N/\Delta t\equiv g$ fixed, and re-exponentiating gives $P_{\text{none}}(R, t)=\exp\left(g\int_{0}^{t}dt'\,\ln S(R, t')\right)$.  Since we are interested in sites sufficiently far from the origin ($R\gg\sqrt{D/z}$) that a given particle has low probability of ever reaching, we can approximate $\ln S=\ln (1-P_{\text{hit}})$ by $-P_{\text{hit}}$, which leads to the following useful expression for $P_{\text{none}}$:
\begin{equation}
P_{\text{none}}(R, t)\simeq\exp\Big[-g\int_{0}^{t}dt'\, (t-t')f_{1}(R, t')\Big]. \label{finalpnone}
\end{equation}
The FPT PDF $f_{N}(R,t)$ is related to $P_{\text{none}}$ by $f_{N}(R,t)=-\partial_{t}P_{\text{none}}(R, t)$.

We will apply these ideas first to a $1$D system on a lattice without convection.  On a lattice, the diffusion constant $D$ is replaced by a hopping rate $w$ (the rate for hopping to a particular side is $w/2$).  We replace the first oasis centered at the origin with a source and ask for the FPT PDF to the $\nu$-th lattice point for the linear model.  By using the proper one-particle FPT PDF $f_{1}(\nu,t)=|\nu|\,e^{-(w+z)t}\,I_{\nu}(wt)/t$ \cite{rednerbook} along with (\ref{finalpnone}), we can find $f_{N}(\nu,t)$ and all of its moments.  In order to compare these values to simulations of the nonlinear model, it is essential to fix $g$ to an appropriate value.  To do this, we match the long-distance $t\to\infty$ solutions for the mean particle concentration in the linear model with a source and the nonlinear model.  The linear model has a $t\to\infty$ solution of $c_{\infty,\text{lin}}(\nu)=g\,e^{-f|\nu|}/w\sinh(f)$, where $f=\cosh^{-1}(1+z/w)$, while the nonlinear model decays like $c_{\infty,\text{nlin}}(\nu)\simeq c_{0}\,e^{-f|\nu|}$.  We have determined $c_{0}$ and thus $g$ numerically.

For the $d=1$ lattice case, $f_{N}(\nu,t)$ is given by a complicated expression, but it does have a simple asymptotic behavior, decaying exponentially like $e^{-\mu(\nu)t}$ as $t\to\infty$, with $\mu(\nu)=g\,e^{-f|\nu|}$.  The $j$-th moment of $f_{N}(\nu, t)$---which we denote $\langle T^{j}(\nu)\rangle$---approaches a simple asymptotic limit as $\nu\to\infty$ \cite{ourlongpaper}:
\begin{equation}
\langle T^{j}(\nu)\rangle=j!\,\frac{e^{f|\nu|j}}{g^{j}}.\quad{(1\text{D lattice})}\label{momlatt}
\end{equation}
The first moment of $f_{N}(\nu,t)$---the \emph{mean first passage time}---thus depends exponentially on the separation of the oases in the limit of large oasis separation.

\begin{figure}
\includegraphics[scale=0.65]{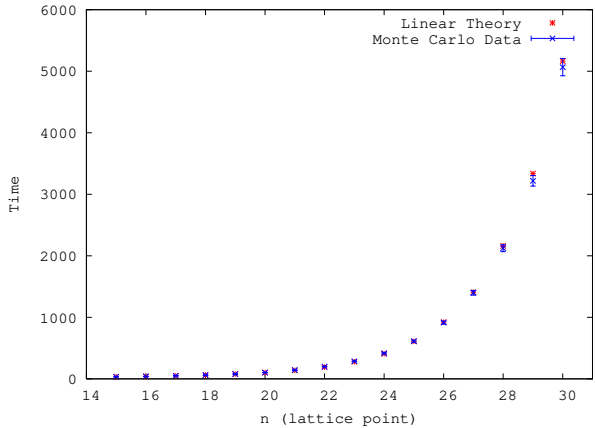}
\caption{\label{MFPT}Comparison of theoretical mean FPT with Monte Carlo data.  The error bars represent a $95\%$ confidence interval.  The parameters used were $w=1.0$, $y=.25$, $z=.1$, and $b=.001$.  Time is measured in units of $1/w$.}
\end{figure}

\begin{figure}
\includegraphics[scale=0.65]{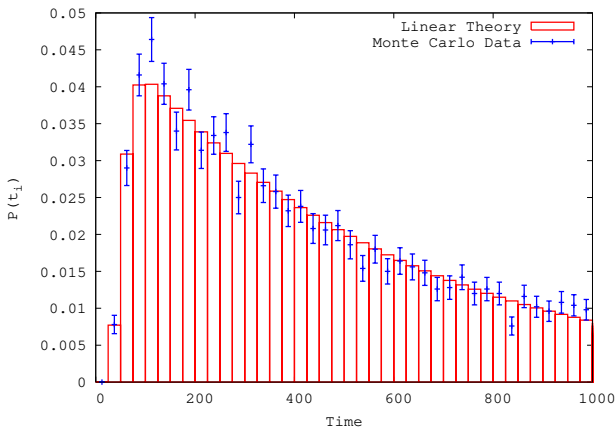}
\caption{\label{histo}Comparison of theory with Monte Carlo data.  Each box represents the probability $P(t_{i})$ (from the linear theory) that a particular run hits site $n=25$ for the first time in the $i$-th time bin.  The points with error bars are from $5000$ Monte Carlo simulations of the full nonlinear stochastic model.  Time is measured as in Fig.\ \ref{MFPT}, and the bins have a width of $25$ units of time.}
\end{figure}

To test the predictions made using the linear model with a source, we wrote a kinetic Monte Carlo program to simulate the full discrete stochastic process in one dimension.  The agreement between the Monte Carlo results and the linear theory with a source is excellent.  The linear theory correctly predicts the mean FPT for distances sufficiently far from the oasis, as can be seen in Fig.\ \ref{MFPT}.  An even more stringent test of the power of the linear theory is a comparison of its prediction of the full FPT PDF with simulation; this is shown in Fig.\ \ref{histo}.  Even with only $5000$ runs, the simulation data begins to fill out the shape of the FPT PDF predicted by the linear theory with a source.

With $g$ specified, the only remaining source of ambiguity in matching up the predictions of the linear theory with a source to the full nonlinear simulation results is the choice of simulation initial conditions.  This ambiguity stems from the fact that the early time dynamics of the linear theory and the nonlinear theory are different; they take different times to ``grow up'' to the point where their mean particle fluxes into the desert are equal.  This disparity becomes less important as transit to sites further away is considered.  For the simulations we performed, we chose to place $y/2b=125$ particles (half the carrying capacity) at the center site of the oasis of $5$ total sites at $t=0$, with no particles at any other site.  This choice led to first passage times in agreement with the linear theory for the parameter sets considered.

The continuum case in one dimension is actually simpler to deal with than the lattice case previously described, even when a uniform convection velocity $v$ is included.  (Physically, $v$ is an external influence on the system, like the motion of a liquid in which bacteria live.)  The FPT PDF $f_{N}(R,t)$ again has a very complicated form, but its behavior in certain limits is easily described.  For any $R$, the function decays exponentially as $e^{-\gamma(R)t}$ as $t\to\infty$, with $\gamma(R)=g(v)\,e^{-\alpha(R)}$ and $\alpha(R)=\sqrt{1+v^2/4Dz}\,\sqrt{z/D}\,R-vR/2D$.  The moments of this distribution are given in the large $R$ limit by \cite{ourlongpaper}
\begin{equation}
\langle T^{j}(R)\rangle=j!\,\frac{e^{\alpha(R)j}}{g(v)^{j}}.\quad{(1\text{D continuum})}
\end{equation}

For higher dimensions ($d>1$), the lattice problem becomes quite complicated, while the continuum problem is still analytically tractable.  Instead of calculating the single particle FPT PDF $f_{1}(R,t)$ to a point, one instead finds $f_{1}(R,a,t)$ to a hypersphere of radius $a$, given an initial condition in which the particle is located at position $\bm{R}$ a distance $|\bm{R}|=R>a$ from the center of the hypersphere \cite{rednerbook}.  As in one dimension, this function is then used along with (\ref{finalpnone}) to determine $f_{N}(R,a,t)$.  For zero convection velocity, we have obtained the following result for the moments of the FPT PDF as $R\to\infty$ \cite{ourlongpaper}: 
\begin{equation}
\langle T^{j}(R,a)\rangle=j!\,\left(\frac{R}{a}\right)^{(d/2-1)j}\,\left(\frac{K_{d/2-1}(\kappa a)}{gK_{d/2-1}(\kappa R)}\right)^j,\label{momcontarbD}
\end{equation}
where $K_{\mu}$ is the $\mu$-th order modified Bessel function of the second kind.  As in the $d=1$ continuum case, it is possible to include the effects of convection.  For a small uniform convection velocity $\bm{v}$, $|\bm{v}|\ll\sqrt{Dz}$, the $j$-th moment is multiplied to leading order by a factor $\left[g(\bm{v}=0)e^{\bm{v}\cdot\bm{R}/2D}/g(\bm{v})\right]^{j}$.  Previous studies concentrating on (\ref{fishnels}) suggest that $g(\bm{v})$ should decrease as $|\bm{v}|$ increases \cite{nelsonshnerb1998}.

We have shown that a linear model with a source can capture the long-distance behavior of the FPT PDF for a nonlinear model with two oases.  We wish to apply our results to a system with many oases.  Consider a very large $d$-dimensional ($d>1$) continuum system in which identical oases of radius $a$ and birth rate $y$ are centered (with overlaps allowed) around randomly placed points with number density $n$ in a desert of death rate $z$.  (Grassberger has studied the related problem of randomly placed traps in a neutral background \cite{grassbergertraps}, and Redner has studied lattice systems with mixed oases and traps \cite{redner84pra}.)  We are interested in the low oasis density regime in which the average distance between oases is much larger than $R_{\text{lin}}$.  One oasis is populated at $t=0$; as time goes on, the population will spread , and eventually an oasis located at position $\bm{L}$ will be reached.  We wish to find the infection time $T_{\text{infection}}$---the time for this process to occur.

Because of the exponential dependence of $K_{d/2-1}(\kappa R)$ on $R$ for large $R$ in (\ref{momcontarbD}), there is a wide distribution of mean transit times between oases in the system.  The situation is mathematically analogous to (though physically quite distinct from) hopping conduction in doped semiconductors: the oases are the analogs of impurity sites, and the mean transit time plays the role of the resistance between sites, which is roughly equal to $e^{\alpha R}/G_{0}$, where $G_{0}$ is a constant.  In the hopping conduction problem, the resistance of the system is dominated by the pairs of impurity sites with the largest separations in the cluster (the critical subnetwork) which carries the bulk of the current \cite{ambegaokarperc,shklovskii}.  This cluster has correlation length $L_{0}$, and above this length scale the system can be regarded as roughly homogenous.  The resistivity $\rho$ is then given by $\rho\simeq{L_{0}}^{d-2}e^{\alpha R_{\text{max}}}/G_{0}$, where $R_{\text{max}}$ is the largest distance between impurity sites in the critical subnetwork, which is roughly equal to the critical percolation radius $R_{c}$ obtained from continuum percolation theory \cite{shklovskii}.  It should be noted that there is some debate as to whether $L_{0}$ is the proper distance to use in $d=3$ for the typical separation of large resistances; one competing theory identifies an additional length scale $l$ which enters into the resistivity along with $L_{0}$ \cite{hunt2005ptf}.

In our problem, the largest oasis separations should dominate the transit time statistics across a sample of size $L_{0}$.  This suggests a coarse-graining that can be used to determine the infection time: we replace each block of size ${L_{0}}^{d}$ with a node, and assign to each node a random transit time picked from $f_{N}(R_{\text{max}},a,t)$, the FPT PDF for crossing the largest oasis separation in each node.  This is a rough approximation; the important point is that the mean time to cross each block is largely determined by the largest oasis separations.  We ignore block-to-block variations in the size of the largest oasis separation, since these become small at the length scale $L_{0}$ \cite{kurkijarvi74}.  The infection time is now given by the path with the shortest total transit time that goes from the block/node containing the starting oasis to the block/node containing the target oasis; we have turned our problem into a first passage percolation (FPP) problem \cite{kesten1986afp}.  It has been shown that, as the distance between nodes goes to infinity, $T_{\text{infection}}/m$, where $m$ is the distance between starting and target nodes, approaches a constant $\mu$, conventionally called the time constant \cite{kesten1986afp}.  An upper limit for $\mu$ is given by the mean time to cross one node.  In the approximation we have made, this is $\langle T(R_{\text{max}},a)\rangle$.  Thus, an estimate for the mean infection time to a distant target site at $\bm{L}$ divided by $|\bm{L}|$ is:
\begin{equation}
\frac{\langle T_{\text{infection}}\rangle}{|\bm{L}|}\simeq\frac{\langle T(R_{\text{max}},a)\rangle}{L_{0}},\label{timeconst}
\end{equation}
where $R_{\text{max}}$ is given by continuum percolation theory as $R_{\text{max}}=[B_{c}(d)/(nV_{d})]^{1/d}$, where $V_{d}$ is the volume of a $d$-dimensional unit hypersphere and $B_{c}(d)$ is a dimensionless number known as the bonding criterion.  The value of $\langle T(R_{\text{max}},a)\rangle$ is given approximately by (\ref{momcontarbD}) for large $R$.  $L_{0}$ is roughly equal to $[\sqrt{z/D}\,R_{\text{max}}]^{\nu}/n^{1/d}$, where $\nu$ is a critical exponent ($\simeq4/3$ in $d=2$, $.88$ in $d=3$) \cite{shklovskii}.

We have presented both analytical and numerical results of investigations of transport in disordered reaction-diffusion systems.  Taking discreteness effects into account, we have shown that the first passage time between two oases separated by a distance $R$ in a desert is a very broadly distributed quantity whose mean increases exponentially with $R$ for large $R$ (see Eq.\ \ref{momcontarbD}).  We have used an analogy with hopping conduction to argue that the largest oasis separations dominate transit times up to a length scale $L_{0}$, and we have employed a mapping to a first passage percolation system to then arrive at an estimate (\ref{timeconst}) for $\langle T_{\text{infection}}\rangle/|\bm{L}|$, the mean infection time to a site located at $\bm{L}$ divided by $|\bm{L}|$.  We are currently working on determining the effects of a uniform convection velocity on our model as well as the nature of the advancing front for the case where half of space, rather than one oasis, is initially populated.  

\begin{acknowledgments}
We would like to thank John Gergely, Richard Sowers, and Uwe T\"auber for helpful discussions, and David Nelson and Nadav Shnerb for help with the numerics.  The simulations were run on the Turing Cluster at UIUC.  This work was supported in part by NSF-DMR grants 03-14279 and 03-25939 (ITR).  
\end{acknowledgments}


\begin{thebibliography}{23}
\expandafter\ifx\csname natexlab\endcsname\relax\def\natexlab#1{#1}\fi
\expandafter\ifx\csname bibnamefont\endcsname\relax
  \def\bibnamefont#1{#1}\fi
\expandafter\ifx\csname bibfnamefont\endcsname\relax
  \def\bibfnamefont#1{#1}\fi
\expandafter\ifx\csname citenamefont\endcsname\relax
  \def\citenamefont#1{#1}\fi
\expandafter\ifx\csname url\endcsname\relax
  \def\url#1{\texttt{#1}}\fi
\expandafter\ifx\csname urlprefix\endcsname\relax\def\urlprefix{URL }\fi
\providecommand{\bibinfo}[2]{#2}
\providecommand{\eprint}[2][]{\url{#2}}

\bibitem[{\citenamefont{Murray}(1993)}]{murray}
\bibinfo{author}{\bibfnamefont{J.~D.} \bibnamefont{Murray}},
  \emph{\bibinfo{title}{Mathematical Biology}}
  (\bibinfo{publisher}{Springer-Verlag}, \bibinfo{address}{New York},
  \bibinfo{year}{1993}).

\bibitem[{\citenamefont{Amar and Family}(1990)}]{isingglauber}
\bibinfo{author}{\bibfnamefont{J.~G.} \bibnamefont{Amar}} \bibnamefont{and}
  \bibinfo{author}{\bibfnamefont{F.}~\bibnamefont{Family}},
  \bibinfo{journal}{Phys.\ Rev.\ A} \textbf{\bibinfo{volume}{41}},
  \bibinfo{pages}{3258} (\bibinfo{year}{1990}).

\bibitem[{\citenamefont{Kroon et~al.}(1993)\citenamefont{Kroon, Fleurent, and
  Sprik}}]{kroon1993}
\bibinfo{author}{\bibfnamefont{R.}~\bibnamefont{Kroon}},
  \bibinfo{author}{\bibfnamefont{H.}~\bibnamefont{Fleurent}}, \bibnamefont{and}
  \bibinfo{author}{\bibfnamefont{R.}~\bibnamefont{Sprik}},
  \bibinfo{journal}{Phys.\ Rev.\ E} \textbf{\bibinfo{volume}{47}},
  \bibinfo{pages}{2462} (\bibinfo{year}{1993}).

\bibitem[{\citenamefont{Birch et~al.}(2007)\citenamefont{Birch, Tsang, and
  Young}}]{birch2007bbf}
\bibinfo{author}{\bibfnamefont{D.}~\bibnamefont{Birch}},
  \bibinfo{author}{\bibfnamefont{Y.}~\bibnamefont{Tsang}}, \bibnamefont{and}
  \bibinfo{author}{\bibfnamefont{W.}~\bibnamefont{Young}},
  \bibinfo{journal}{Phys. Rev. E} \textbf{\bibinfo{volume}{75}},
  \bibinfo{pages}{66304} (\bibinfo{year}{2007}).

\bibitem[{\citenamefont{Janssen}(1997)}]{runawayRG}
\bibinfo{author}{\bibfnamefont{H.~K.} \bibnamefont{Janssen}},
  \bibinfo{journal}{Phys.\ Rev.\ E} \textbf{\bibinfo{volume}{55}},
  \bibinfo{pages}{6253} (\bibinfo{year}{1997}).

\bibitem[{\citenamefont{Moreira and Dickman}(1996)}]{MCofcontactproc}
\bibinfo{author}{\bibfnamefont{A.~G.} \bibnamefont{Moreira}} \bibnamefont{and}
  \bibinfo{author}{\bibfnamefont{R.}~\bibnamefont{Dickman}},
  \bibinfo{journal}{Phys. Rev. E} \textbf{\bibinfo{volume}{54}},
  \bibinfo{pages}{R3090} (\bibinfo{year}{1996}).

\bibitem[{\citenamefont{Vojta}(2004)}]{disorderPTVojta}
\bibinfo{author}{\bibfnamefont{T.}~\bibnamefont{Vojta}},
  \bibinfo{journal}{Phys.\ Rev.\ E} \textbf{\bibinfo{volume}{70}},
  \bibinfo{eid}{026108} (\bibinfo{year}{2004}).

\bibitem[{\citenamefont{Joo and Lebowitz}(2005)}]{jaewookcontact}
\bibinfo{author}{\bibfnamefont{J.}~\bibnamefont{Joo}} \bibnamefont{and}
  \bibinfo{author}{\bibfnamefont{J.~L.} \bibnamefont{Lebowitz}},
  \bibinfo{journal}{Phys. Rev. E} \textbf{\bibinfo{volume}{72}},
  \bibinfo{pages}{036112} (\bibinfo{year}{2005}).

\bibitem[{\citenamefont{Park and Deem}(1998)}]{parkanddeem}
\bibinfo{author}{\bibfnamefont{J.-M.} \bibnamefont{Park}} \bibnamefont{and}
  \bibinfo{author}{\bibfnamefont{M.~W.} \bibnamefont{Deem}},
  \bibinfo{journal}{Phys. Rev. E} \textbf{\bibinfo{volume}{57}},
  \bibinfo{pages}{3618} (\bibinfo{year}{1998}).

\bibitem[{\citenamefont{Nelson and Shnerb}(1998)}]{nelsonshnerb1998}
\bibinfo{author}{\bibfnamefont{D.~R.} \bibnamefont{Nelson}} \bibnamefont{and}
  \bibinfo{author}{\bibfnamefont{N.~M.} \bibnamefont{Shnerb}},
  \bibinfo{journal}{Phys. Rev. E} \textbf{\bibinfo{volume}{58}},
  \bibinfo{pages}{1383} (\bibinfo{year}{1998});
\bibinfo{author}{\bibfnamefont{K.~A.} \bibnamefont{Dahmen}},
  \bibinfo{author}{\bibfnamefont{D.~R.} \bibnamefont{Nelson}},
  \bibnamefont{and} \bibinfo{author}{\bibfnamefont{N.~M.} \bibnamefont{Shnerb}}
  (\bibinfo{year}{1999}), \urlprefix\url{arXiv.org:cond-mat/9903276};
\bibinfo{author}{\bibfnamefont{K.~A.} \bibnamefont{Dahmen}},
  \bibinfo{author}{\bibfnamefont{D.~R.} \bibnamefont{Nelson}},
  \bibnamefont{and} \bibinfo{author}{\bibfnamefont{N.~M.}
  \bibnamefont{Shnerb}}, \bibinfo{journal}{J.\ Math.\ Biol.}
  \textbf{\bibinfo{volume}{41}}, \bibinfo{pages}{1} (\bibinfo{year}{2000}).

\bibitem[{\citenamefont{Lin et~al.}(2004)\citenamefont{Lin, Mann,
  Torres-Oviedo, Lincoln, Kas, and Swinney}}]{exper2}
\bibinfo{author}{\bibfnamefont{A.~L.} \bibnamefont{Lin}}
  \bibinfo{author}{\bibfnamefont{{\it et.}} \bibnamefont{{\it al}}},
  \bibinfo{journal}{Biophys. J.} \textbf{\bibinfo{volume}{87}},
  \bibinfo{pages}{75} (\bibinfo{year}{2004}).

\bibitem[{\citenamefont{M\'endez et~al.}(2003)\citenamefont{M\'endez, Fort,
  Rotstein, and Fedotov}}]{fisherwavesrandommedia}
\bibinfo{author}{\bibfnamefont{V.}~\bibnamefont{M\'endez}}
  \bibinfo{author}{\bibfnamefont{{\it et.}}~\bibnamefont{{\it al}}},
  \bibinfo{journal}{Phys. Rev. E} \textbf{\bibinfo{volume}{68}},
  \bibinfo{pages}{041105} (\bibinfo{year}{2003}).

\bibitem[{\citenamefont{Armero et~al.}(1996)\citenamefont{Armero, Sancho,
  Casademunt, Lacasta, Ram\'irez-Piscina, and Sagu\'es}}]{flucsinfronts}
\bibinfo{author}{\bibfnamefont{J.}~\bibnamefont{Armero}}
  \bibinfo{author}{\bibfnamefont{{\it et.}} \bibnamefont{{\it al}}},
  \bibinfo{journal}{Phys. Rev. Lett.} \textbf{\bibinfo{volume}{76}},
  \bibinfo{pages}{3045} (\bibinfo{year}{1996});
\bibinfo{author}{\bibfnamefont{J.}~\bibnamefont{Armero}}
  \bibinfo{author}{\bibfnamefont{{\it et.}}~\bibnamefont{{\it al}}},
\bibinfo{journal}{Phys. Rev. E}
  \textbf{\bibinfo{volume}{58}}, \bibinfo{pages}{5494} (\bibinfo{year}{1998}).

\bibitem[{\citenamefont{Santos and Sancho}(1999)}]{noisefronts}
\bibinfo{author}{\bibfnamefont{M.~A.} \bibnamefont{Santos}} \bibnamefont{and}
  \bibinfo{author}{\bibfnamefont{J.~M.} \bibnamefont{Sancho}},
  \bibinfo{journal}{Phys. Rev. E} \textbf{\bibinfo{volume}{59}},
  \bibinfo{pages}{98} (\bibinfo{year}{1999}).

\bibitem[{\citenamefont{Redner}(2001)}]{rednerbook}
\bibinfo{author}{\bibfnamefont{S.}~\bibnamefont{Redner}},
  \emph{\bibinfo{title}{A Guide to First-Passage Processes}}
  (\bibinfo{publisher}{Cambridge University Press},
  \bibinfo{address}{Cambridge, UK}, \bibinfo{year}{2001}).

\bibitem[{\citenamefont{Dahmen and Missel}()}]{ourlongpaper}
\bibinfo{author}{\bibfnamefont{K.~A.} \bibnamefont{Dahmen}} \bibnamefont{and}
  \bibinfo{author}{\bibfnamefont{A.~R.} \bibnamefont{Missel}},
  \bibinfo{note}{unpublished}.

\bibitem[{\citenamefont{Grassberger and Procaccia}(1982)}]{grassbergertraps}
\bibinfo{author}{\bibfnamefont{P.}~\bibnamefont{Grassberger}} \bibnamefont{and}
  \bibinfo{author}{\bibfnamefont{I.}~\bibnamefont{Procaccia}},
  \bibinfo{journal}{Phys. Rev. A} \textbf{\bibinfo{volume}{26}},
  \bibinfo{pages}{3686} (\bibinfo{year}{1982}).

\bibitem[{\citenamefont{Redner and Kang}(1984)}]{redner84pra}
\bibinfo{author}{\bibfnamefont{S.}~\bibnamefont{Redner}} \bibnamefont{and}
  \bibinfo{author}{\bibfnamefont{K.}~\bibnamefont{Kang}},
  \bibinfo{journal}{Phys. Rev. A} \textbf{\bibinfo{volume}{30}},
  \bibinfo{pages}{3362} (\bibinfo{year}{1984}).

\bibitem[{\citenamefont{Shklovskii and Efros}(1984)}]{shklovskii}
\bibinfo{author}{\bibfnamefont{B.~I.} \bibnamefont{Shklovskii}}
  \bibnamefont{and} \bibinfo{author}{\bibfnamefont{A.~L.} \bibnamefont{Efros}},
  \emph{\bibinfo{title}{Electronic Properties of Doped Semiconductors}}
  (\bibinfo{publisher}{Springer-Verlag}, \bibinfo{address}{Berlin},
  \bibinfo{year}{1984}).

\bibitem[{\citenamefont{Ambegaokar et~al.}(1971)\citenamefont{Ambegaokar,
  Halperin, and Langer}}]{ambegaokarperc}
\bibinfo{author}{\bibfnamefont{V.}~\bibnamefont{Ambegaokar}},
  \bibinfo{author}{\bibfnamefont{B.~I.} \bibnamefont{Halperin}},
  \bibnamefont{and} \bibinfo{author}{\bibfnamefont{J.~S.}
  \bibnamefont{Langer}}, \bibinfo{journal}{Phys. Rev. B}
  \textbf{\bibinfo{volume}{4}}, \bibinfo{pages}{2612} (\bibinfo{year}{1971}).

\bibitem[{\citenamefont{Hunt}(2005)}]{hunt2005ptf}
\bibinfo{author}{\bibfnamefont{A.}~\bibnamefont{Hunt}},
  \emph{\bibinfo{title}{{Percolation Theory for Flow in Porous Media}}}
  (\bibinfo{publisher}{Springer}, \bibinfo{year}{2005}).

\bibitem[{\citenamefont{Kurkij\"arvi}(1974)}]{kurkijarvi74}
\bibinfo{author}{\bibfnamefont{J.}~\bibnamefont{Kurkij\"arvi}},
  \bibinfo{journal}{Phys. Rev. B} \textbf{\bibinfo{volume}{9}},
  \bibinfo{pages}{770} (\bibinfo{year}{1974}).

\bibitem[{\citenamefont{Kesten}(1986)}]{kesten1986afp}
\bibinfo{author}{\bibfnamefont{H.}~\bibnamefont{Kesten}},
  \bibinfo{journal}{Lecture Notes in Math} \textbf{\bibinfo{volume}{1180}},
  \bibinfo{pages}{125} (\bibinfo{year}{1986}).

\end{thebibliography}
\end{document}